# Multi-Compartmental Biomaterial Scaffolds for Patterning Neural Tissue Organoids in Models of Neurodevelopment and Tissue Regeneration

Richard J. McMurtrey

Institute of Neural Regeneration & Tissue Engineering, Highland, UT, United States
Email: *richard.mcmurtrey@neuralregeneration.org*

*Abstract*

Biomaterials are becoming an essential tool in the study and application of stem cell research. Various types of biomaterials enable three-dimensional (3D) culture of stem cells, and, more recently, also enable high-resolution patterning and organization of multicellular architectures. Biomaterials also hold potential to provide many additional advantages over cell transplants alone in regenerative medicine. This paper describes novel designs for functionalized biomaterial constructs that guide tissue development to targeted regional identities and structures. Such designs comprise compartmentalized regions in the biomaterial structure that are functionalized with molecular factors that form concentration gradients through the construct and guide stem cell development, axis patterning, and tissue architecture, including rostral/caudal, ventral/dorsal, or medial/lateral identities of the central nervous system. The ability to recapitulate innate developmental processes in a 3D environment and under specific controlled conditions has vital application to advanced models of neurodevelopment and for repair of specific sites of damaged or diseased neural tissue.

*Introduction*

It has long been recognized that stem cells hold tremendous therapeutic potential for a variety of diseases and injuries, particularly for numerous ailments that have no effective treatments, but restoring tissue and organ functions with stem cells has also posed several challenges that remain to be overcome, including sufficient cell survival, development, and functional integration in native tissues. One approach for surmounting many of these limitations and significantly advancing capabilities of stem cells in therapeutic applications is the use of stem cells cultured in biomaterial architecture. Biomaterials offer numerous advantages and expand capabilities that generally cannot be achieved with stem cells alone, particularly in the case of implanting cells and reconstructing damaged tissue.

Biomaterials generally include natural or synthetic matrix molecules (or a modified combination of the two) that replicate aspects of innate extracellular matrix in various tissues. This biomaterial scaffolding can serve many functions, such as supporting structural organization and patterning of cells, providing molecular and mechanical differentiation cues, arranging attachment points for cells and aiding anchorage-dependent survival, as well as preventing cells from being washed away from implantation sites in the body [1]. Many types of biomaterials have been shown to influence the survival and function of developing stem cells, and biomaterial designs can help replicate cellular interactions, matrix characteristics, biochemical gradients, and signaling events that occur in development, in addition to supporting cell survival, differentiation, and integration into innate tissue.

The suspension of stem cells in biomaterial polymers and aqueous media thus enables the formation of unique structures and functions found in many types of organ tissues. One example of this is cerebral organoids, where clusters of pluripotent stem cells are cultured in spheres of proteinaceous Matrigel to create numerous types of neural structures and cell types [2]. 3D hydrogel cultures of stem cells have been shown to enable the formation of other advanced anatomical structures, including various gastrointestinal, hepatic, pancreatic, renal, retinal, and neural tissues. The cells themselves appear to possess innately-programmed capabilities to self-assemble at least some aspects of important anatomical structures even in unpatterned hydrogel constructs. For example, neural tissue organoids have demonstrated various aspects of ventricular, hippocampal, retinal, spinal, and cortical regions, and such organoid constructs are being used to identify previously unknown mechanisms of neurological diseases like microcephaly, autism, neurodegenerative diseases, zika virus infection, and others [2-6] as well as to investigate normal tissue development [7-9].

Importantly, the use of biomaterials in clinically-relevant testing has also shown promise in enhancing cell survival and cell migration at the implantation site [10-12], promoting a favorable regenerative environment [13], and extending neural connectivity through neural lesions [14]. The presence of hydrogel materials loaded with neurotrophic factors and implanted in ischemic brain tissue can enhance cell survival, axonal sprouting, and migration of immature neuronal cells around the stroke area [12]. Thus engineered tissues hold great potential in regenerative medicine, but one of the weaknesses of stem-cell-derived tissues is the lack of comprehensive control over specific regional identities, architectures, and cellular sub-specializations in the biomaterials. Accurate replication of complex structures like cortex, hippocampus, retina, tracts, or other neuroanatomy cannot yet be entirely controlled or achieved and each organoid may vary significantly in the resulting composition of cell types, structures, and self-organization patterns.

*Synthesis of Form and Function*

Because a simple homogenous biomaterial sphere is likely not sufficient to consistently guide all types of cellular self-organization and tissue patterning in organoids, advancing technologies seek to enable composite biomaterials to provide more detailed guidance of cellular architecture and identity. This may be accomplished through a variety of means. The architecture of a biomaterial construct, for example, can be formed with committed or restricted neural stem cells capable of forming diverse neuroglial subtypes and the inclusion of patterned fiber scaffolds suspended in the biomaterial can guide and enhance cell attachment and neurite outgrowth along the fibers [11, 15]. Hydrogels functionalized with axonal guidance molecules may enhance cellular attachment, migration, differentiation, and axonal extension, and hydrogels seeded with diseased cells can replicate innate pathological features of diseases like Alzheimer's and Parkinson's disease [2-4, 16]. Characteristics like biomaterial stiffness, density, and cross-linking capability can influence cell differentiation and function [17], and a wide array of tissue types and cell states may be achieved with the choice of biomaterial, either by means of intrinsic cell signals or by means of various growth factors and differentiation factors that can be supplied within the biomaterials.

In addition, biomaterials may be designed to enhance the biocompatibility of tissue implants. Certain types of matrix molecules may inhibit inflammatory and scarring reactions that arise from disrupting tissue or introducing a foreign body into the tissue, either by interacting with cellular receptors or by absorbing and diffusing reactive cell signaling factors. If the biomaterial takes too long to degrade or has too great of stiffness compared with the innate tissue, it may trigger foreign body reactions that prevent functional integration of the implanted cells, and conversely, if the biomaterial degrades too quickly or is too soft or friable, it may not adequately support functional integration of implanted cells into the tissue. Thus biomaterial polymers must be optimized to facilitate integration into host tissue, minimize foreign-body reactions, and enhance the permissive environment of the tissue.

*The Role of Diffusion*

Recent research has suggested that the inherent diffusion limitations of gasses, nutrients, and signaling molecules through 3D tissue cultures can specifically affect cellular organization, differentiation state, and metabolic characteristics of cells [18]. While this phenomenon can impede important oxygen and nutrient delivery to cells in a 3D environment, it can also be advantageously used to replicate molecular concentration gradients in synthetic 3D tissue constructs. In other words, the biomaterial can serve to enclose and contain local regions with concentration gradients of certain factors in the construct that may be endogenously secreted by cells or that may be introduced artificially. Such a capability is vital for replicating innate gradients and cues that occur in early tissue development or that may play a role in certain pathological conditions, and by forming compartments of diffusing factors in the biomaterial construct, concentration gradients can be formed to drive stem cell differentiation and guide tissue development. These concentration gradients may be formed of trophic factors, migratory cues, neurite guidance signals, hormones, nutrients, or other factors that affect tissue patterning and cell identity at distinct regions of a tissue construct. Understanding these diffusion signals and processes will better help deliver specific controllable concentrations of ions, nutrients, and other factors to specific regions of the cellular tissue constructs, which in turn will enable tight control of cell state and differentiation processes. Evidence also suggests that stem cells implanted into the body may have better survival if they are cultured and prepared under certain conditions of hypoxic stress or ischemic exposure, a form of "preconditioning" the cells [19], and culturing cells in 3D conditions with limited diffusion is one way of achieving the preconditioning effect for regenerative cell therapy applications.

*3D Biomaterial Scaffolding Designs*

Several examples of 3D biomaterial designs are given herein that comprise distinct nanoarchitecture, functionalized compartments, and microgradients in the tissue construct [Fig. 1]. These designs provide detailed control of cell differentiation and tissue structure, and several examples are described demonstrating how certain innate neurodevelopmental cues can be mimicked in a 3D tissue construct [Fig. 2] and how regional patterning of cortical tissue can be influenced with applied gradients of signaling factors [Fig. 3].

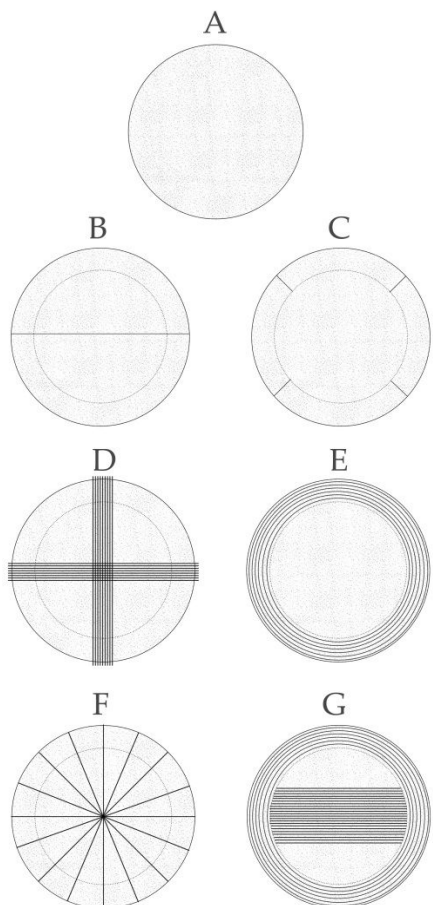

*Figure 1: Examples of various organoid scaffolding designs created with biomaterial hydrogels and polymer fibers. (A) represents the standard organoid matrix droplet in which cellular spheroids are placed for organoid differentiation. (B-G) represent more complex designs with two or more distinct regions or*

*polarized compartments in the biomaterial, such as where an external hydrogel layer is loaded with signaling molecules, growth factors, or functionalized fibers and then coated around the inner hydrogel sphere. Each compartment is used to create certain effects—for example, in (B) the upper compartment may be biochemically functionalized with rostral-inducing factors while the lower compartment may be functionalized with caudal-inducing factors. As another example, in (C) the lower and upper compartments may be functionalized with ventral-dorsal directional gradients and the remaining central and outer compartments may be functionalized in a lateral-medial-lateral manner, as described in the text. The compartments of the hydrogel may be directly loaded with corresponding signaling factors to create concentration gradients, or the fibers or hydrogel polymers themselves may be functionalized with the desired signaling factor. Fibers are about a micron in diameter while the tissue constructs are a few millimeters in diameter, and thus the figures are not to scale.*

Though appearing complicated, these designs may be constructed through relatively simple means. Biochemical signaling factors (which may include differentiation cues, patterning factors, axonal guidance molecules, growth factors, or others) are loaded or mixed into specific compartments of biomaterial hydrogel at desired initial concentrations, and these then diffuse and form specific concentration gradients through the biomaterial construct. These concentration gradient profiles of the signaling factors through the spatial coordinates of the construct over time were recently derived and published, including for scenarios of a limited supply of diffusant molecules (like certain signaling factors or glucose in media) or for an unlimited diffusant supply (like oxygen in culture), and with or without metabolism of the diffusant by cells in the 3D tissue construct [18].

Functionalized nanofibers themselves may also be suspended within or through the 3D hydrogel (as in Figures 1D-G), the method of which has been described previously [15]. Functionalized fibers may also be rolled into the hydrogel (as in Figure 1E), and the biomaterial layers of the construct may be assembled in stages rather than all at once, in order, for example, to add external signaling gradients at later time points in the culture. The nanofibers are functionalized by coating biochemical factors directly onto the polymer fibers. If coating by adsorptive methods, the factors may then diffuse from the fibers into the hydrogel compartments, or if coating by more permanent chemical coupling methods, guidance molecules or attachment domains remain bound to the polymer fibers. By exploring a variety of factors and arrangements, such constructs enable detailed study of what signals and structures are necessary and sufficient (and what signals are influential but non-essential) for neurodevelopmental processes like cell sub-specialization, axis patterning, regionalization, and architecture formation.

Biochemically-active molecules that serve as differentiation and patterning factors can guide tissue construct polarization and axis patterning by being incorporated into specific regions of the hydrogel. Diffusive gradients of these factors can then be established in a similar orientation to innate biochemical gradients that are known to influence developmental processes in central neural tissue. For example, in the brain and spinal cord, ventralization effects and the corresponding neural and astroglial identities may be achieved by incorporating ventralization factors like sonic hedgehog protein (SHH), smoothened agonist (SAG), or purmorphamine into a hydrogel compartment [20-23]. Fibroblast growth factor 8 (FGF-8) can be added to induce midbrain dopaminergic identity [24-26]. Dorsalization effects in the tissue construct may likewise be achieved with "wnt" protein (WNT), bone morphogenetic protein 4 (BMP-4), cyclopamine, or others. WNT actually plays a complex role and may help to induce certain dorsalization, caudalization, or ventralization effects depending on context [22, 27-32]. Similarly, the family of fibroblast growth factors (FGFs) can play varied roles in caudalization, rostralization, ventralization, and dorsalization [33-35]. Other small molecule inhibitors of the above-mentioned pathways can also be used to divert commitment towards opposite fates—for example, while WNT, BMP, and nodal can induce caudal fates, WNT inhibitors, BMP inhibitors, and nodal inhibitors may be implemented to drive a rostral fate. Thus signaling molecule gradients in synthetic forms can be used to recapitulate the natural directional gradients of SHH, WNT, BMP, and FGFs that are essential in proper neural tissue development.

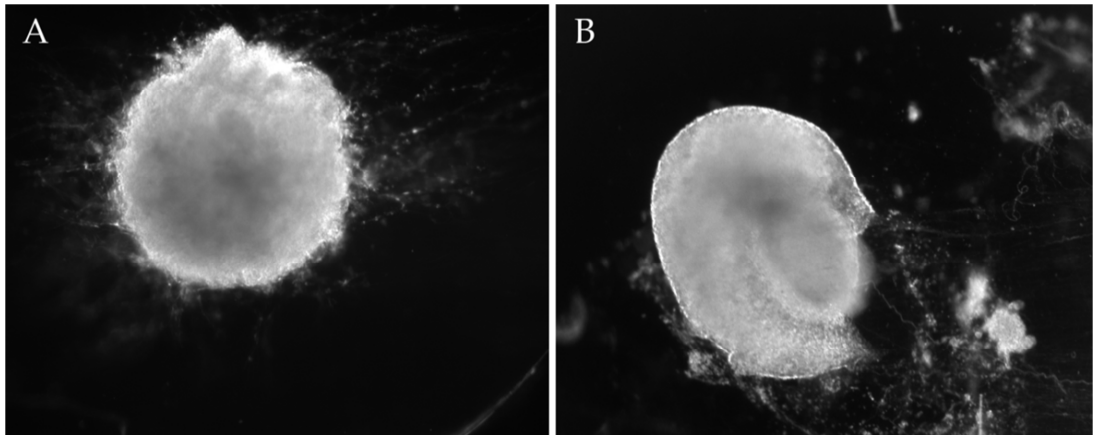

*Figure 2: Functionalized nanofibers enable a greater variety of architectural formations in cerebral organoids grown from induced pluripotent stem cells (iPSCs). (A) Sphere of neurally-induced iPSCs beginning to migrate along functionalized nanofibers in a hydrogel. (B) Hydrogel rolled with functionalized nanofibers resembling a folded cortex similar to the folding of the hippocampal formation.*

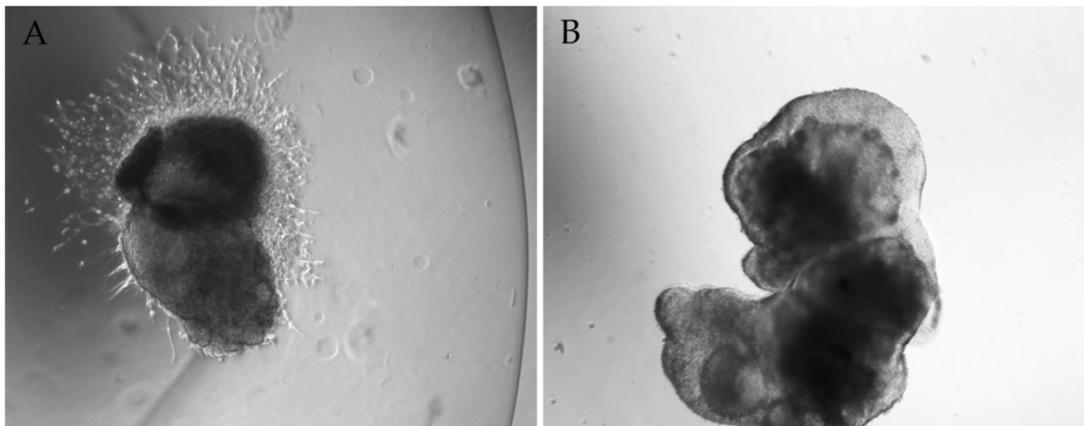

*Figure 3: The choice of biomaterial composition and molecular signaling factors produces different effects on spheroids of differentiating iPSCs. (A) Dual phase spherical hydrogel with neural differentiation and prolific neurite outgrowth. (B) Regionalized neuroepithelial development (seen as the more translucent layer around the tissue) in a biomaterial construct.*

Most regionalization factors, however, cannot simply be partitioned into clear categorical effects since they have interacting effects and depend substantially on timing, concentration, and cell state. For example, over decades of research work on neurodevelopment, it was found that endogenous inhibition of the BMP molecule was required for the formation of neural cells from the early ectodermal layer in many developing species from amphibians to mammals [36-37]. Through the 1990s and early 2000s it was found that inhibitors of BMP signaling (such as dorsomophin, noggin, chordin, and follistatin) as well as inhibitors of TGF-β (such as SB-431542) could be used in the culture of stem cells to promote neural differentiation. Further research suggested that these signaling pathways, while necessary for neural induction, were not entirely sufficient. More complex roles were found for FGF signaling, WNT signaling, and retinoic acid (RA) signaling, each of which could influence neural differentiation based on the timing and concentration of signaling factors [2, 38-40]. FGF signaling has some downstream effects on inhibiting BMP and SMAD family member proteins, but FGF also produces a neural fate independently of BMP or SMAD modulation [38]. FGF appears to have numerous effects, including promoting neural induction, promoting proliferation of neural stem cells, promoting a caudal fate, and, if the cells have already committed to a cerebral fate, promoting a ventral forebrain identity [41].

Retinoic acid can also induce neuroglial lineage in the brain [42], with high concentrations tending to produce neurons that are representative of the caudal portion of the brain and spinal cord [41, 43-44] (similar to the caudalizing effects of FGF-2), although again its effects depend on concentration, timing, and current state of the cells. Early exposure of neural progenitor cells to high levels of RA tends to promote posteriorization, while activation of SHH signaling promotes ventralization. It has been found that RA can promote SHH activity, and gradients of both RA and SHH are important for the induction of the 'floor plate,' a tissue structure at the front of the neural tube that helps signal ventral-dorsal patterning of the cells within the cord. In the spinal cord, these factors are influential in driving ventral motor neuron identity and central interneuron identity [7]. RA also interacts with WNT by altering gene promoter assembly and activating alternate, non-canonical WNT pathways (such as differentiation signals) rather than canonical WNT pathways (such as maintenance of stemness) [45].

During development the meningeal layers that surround the brain play a significant role in neurodevelopment by providing retinoic acid signaling and chemoattractant signals that aid outer attachment of radial glial fibers and by promoting the migration of neural cells to the cortex [2, 46-47]. The dura, which is the most outer layer of the meninges, also secretes FGF-2 and TGF-β related molecules that aid in the initiation of bone plates that become the skull. As neural stem cells migrate from central progenitor regions along spoke-like radial glia to populate the outer rims that will become cortex, network connections both near and far also begin to form.

Also early in development of the brain, reelin is expressed and secreted from Cajal-Retzius neurons, which are the primary cells found in the outer layer I of the early-developing cortex. It is thought that reelin helps guide radial glia fibers to extend outwardly towards the edges of the brain where cortex and hippocampus will eventually be formed [48-53], and reelin enables neurons and neural progenitors to release from radial glia fibers and migrate throughout the cortex [51, 54-55]. It has also been shown that reelin is essential for neuronal positioning in cortex and spinal cord [51, 56], where cortical and hippocampal fibers orient along the reelin gradient [48, 53]. A lack of reelin causes lissencephaly or agyria, likely due to the fact that radial glia scaffolding that neurons rely on for migration to the cortex is deformed and cannot provide the proper guide for neural precursors [49-50, 52].

Reelin has also been implicated in diseases like Alzheimer's, epilepsy, schizophrenia, and other disorders [57-61]. The mechanisms involved are varied, but appear to involve interactions among reelin-responsive domains in apolipoprotein E receptors and very low-density lipoprotein receptors, which have been shown to be involved in radial migration of neuronal precursors and detachment of such precursors from migratory chains [62]. Reelin also can promote dendritic growth and strengthen long-term potentiation by influencing N-methyl-D-aspartate (NMDA) receptors [63-68], and it is thought that reelin increases the mobilization of the NMDA receptors such that they "float" among the lipid molecules of the membrane rather than being anchored to the architecture of the cell. The ability to influence membrane mobility allows reelin to control subunit expression within NMDA receptors after early development [69], and this may allow the NMDA receptor to mature more quickly, since the maturation of the NMDA receptor involves replacing the NR2B subunit with the NR2A subunit [70].

As brain development proceeds, signals that control area specification continue in complexity, and the cells of the brain not only possess the machinery for all the extraordinary functions of thought and memory, but also the incredible capability to self-organize and assemble its complex pathways and networks. Numerous transcription factors determine finely detailed patterns of brain development and spatial identity. One of the more simple examples from this complexity is found in the dueling roles of Emx2 and Pax6. In the 3D environment of the brain, Emx2 directs the development of the caudal-medial region while Pax6 directs the rostral-lateral region of the cortex. Each of these factors inhibits the other, thereby making a strong gradient of Pax6 in one region and a strong gradient of Emx2 in the opposite region. If either one of these factors is inactive or mutated, the cortical region it is responsible for will be underdeveloped and the opposing region will overgrow [71-73]. Expression of both Pax6 and Emx2 is strongly influenced by concentrations of previously mentioned factors like SHH (which represses Pax6) and FGFs (like FGF-8 which represses Emx2) [33, 74-75], meaning that although simple regional gradients can be generated in organoid constructs, additional downstream gradients of signaling factors may complicate or interweave with intended results.

As can be seen, investigating the function of just a single type of diffusing signaling factor in the brain can be quite complex, meaning that assembling the roles of thousands of possible variations is a daunting task, but the advent of these 3D tissue designs enables a platform for controlled study of individual factors in neural development and disease. Recent work suggests that the cortex has at least 180 distinct and conserved areas [76], the identities and structures of which must be shaped from a variety of signaling cues. The ability to replicate these innate neurodevelopmental processes and three-dimensional compositions in the lab with known gradients and localized concentrations of molecular signaling factors is essential for directed specification and accurate study of stem cell development and thus vital to the targeted regeneration of functional neural tissue.

*Conclusions*

When attempting to design ideal cellular constructs for implantation, numerous aspects must be carefully considered in the context of the specific tissue type and the desired functional effects. Restoring function in tissues and organs—whether brain, heart, lung, liver, kidney, nerve, muscle, or other tissues—is likely to require integrating and reconstructing functional architecture of the tissue rather than simply implanting isolated cells. Recent stem cell research has therefore focused on the application of biomaterials along with stem cells to enable desired patterning and organization of more accurate and complete tissue structures. The tissue construct designs presented herein enable an array of important investigations in tissue development, disease mechanisms, environmental toxicologies, and regenerative medicine applications.

Nevertheless, much research remains to be done on the optimal combinations of biomaterials, signaling factors, and scaffolding architectures needed to optimally prepare cells for transplantation and integration into specific tissues of the body. This work describes novel biomaterial designs for controlling critical processes of neurodevelopment, within which molecular signaling gradients may be established according to physical diffusion models. It is not yet known to what extent neural networks can be functionally established in 3D tissue constructs, or whether 3D architecture can be integrated into the prohibitive *in vivo* central nervous system environment, but the ability to accurately replicate neuroanatomical structures and guide neural network formation in three dimensions, both *in vitro* and *in vivo*, will produce major advances in the field of neural regeneration.

*Disclosure Statement*

The author declares no potential conflicts of interest with respect to the research, authorship, and/or publication of this article and was not granted financial support for this work.